\documentclass[]{spie}
\usepackage[]{graphicx}
\usepackage[usenames]{color}
\usepackage{hyperref}
\usepackage[utf8x]{inputenc}

\title{ASTRA: ASTrometry and phase-Referencing Astronomy\\
on the Keck interferometer} 

\author{J. Woillez\supit{a}, R. Akeson\supit{b}, M. Colavita\supit{c}, J. Eisner\supit{d}, A. Ghez\supit{e}, J. Graham\supit{f}, L. Hillenbrand\supit{g}, R. Millan-Gabet\supit{b}, J. Monnier\supit{h}, J.-U. Pott\supit{i}, S. Ragland\supit{a}, P. Wizinowich\supit{a},
\skiplinehalf
E. Appleby\supit{a}, B. Berkey\supit{a}, A. Cooper\supit{a}, C. Felizardo\supit{b}, J. Herstein\supit{b}, M. Hrynevych\supit{a},\\
O. Martin\supit{a}, D. Medeiros\supit{a}, D. Morrison\supit{a}, T. Panteleeva\supit{a}, B. Smith\supit{a},\\
K. Summers\supit{a}, K. Tsubota\supit{a}, C. Tyau\supit{a}, E. Wetherell\supit{a}
\skiplinehalf
{\small	\supit{a}W. M. Keck Observatory, 65-1120 Mamalahoa Hwy, Kamuela, HI 96743, USA; \\
		\supit{b}NExScI, California Institute of Technology, 770 South Wilson Ave, Pasadena, CA 91125, USA; \\
		\supit{c}JPL, California Institute of Technology, 4800 Oak Grove Drive, Pasadena, CA 91109, USA; \\
		\supit{d}University of Arizona, 933 North Cherry Avenue, Tucson, AZ 85721-0065, USA; \\
		\supit{e}University of California Los Angeles, Los Angeles, CA 90095-1547, USA; \\
		\supit{f}University of California Berkeley, 601 Campbell Hall, Berkeley, CA 94720-3411, USA; \\
		\supit{g}California Institute of Technology, Pasadena, CA 91125, USA; \\
		\supit{h}University of Michigan, 941 Dennison Bldg, AnnArbor, MI 48109-1090, USA; \\
		\supit{i}Max-Planck-Institut f\"ur Astronomie, K\"onigstuhl 17, D-69117, Heidelberg, Germany. \\}
}

\authorinfo{Send correspondence to J. Woillez - E-mail: jwoillez@keck.hawaii.edu - Telephone: +1 (808) 881-35114}
 
\begin{document} 
  \maketitle 

\begin{abstract}
ASTRA (ASTrometric and phase-Referencing Astronomy) is an upgrade to the existing Keck Interferometer which aims at providing new self-phase referencing (high spectral resolution observation of YSOs), dual-field phase referencing (sensitive AGN observations), and astrometric (known exoplanetary systems characterization and galactic center general relativity in strong field regime) capabilities. With the first high spectral resolution mode now offered to the community, this contribution focuses on the progress of the dual field and astrometric modes.
\end{abstract}



\section{INTRODUCTION}
\label{sec:intro}

\begin{figure}
\centering
\includegraphics[width=\linewidth]{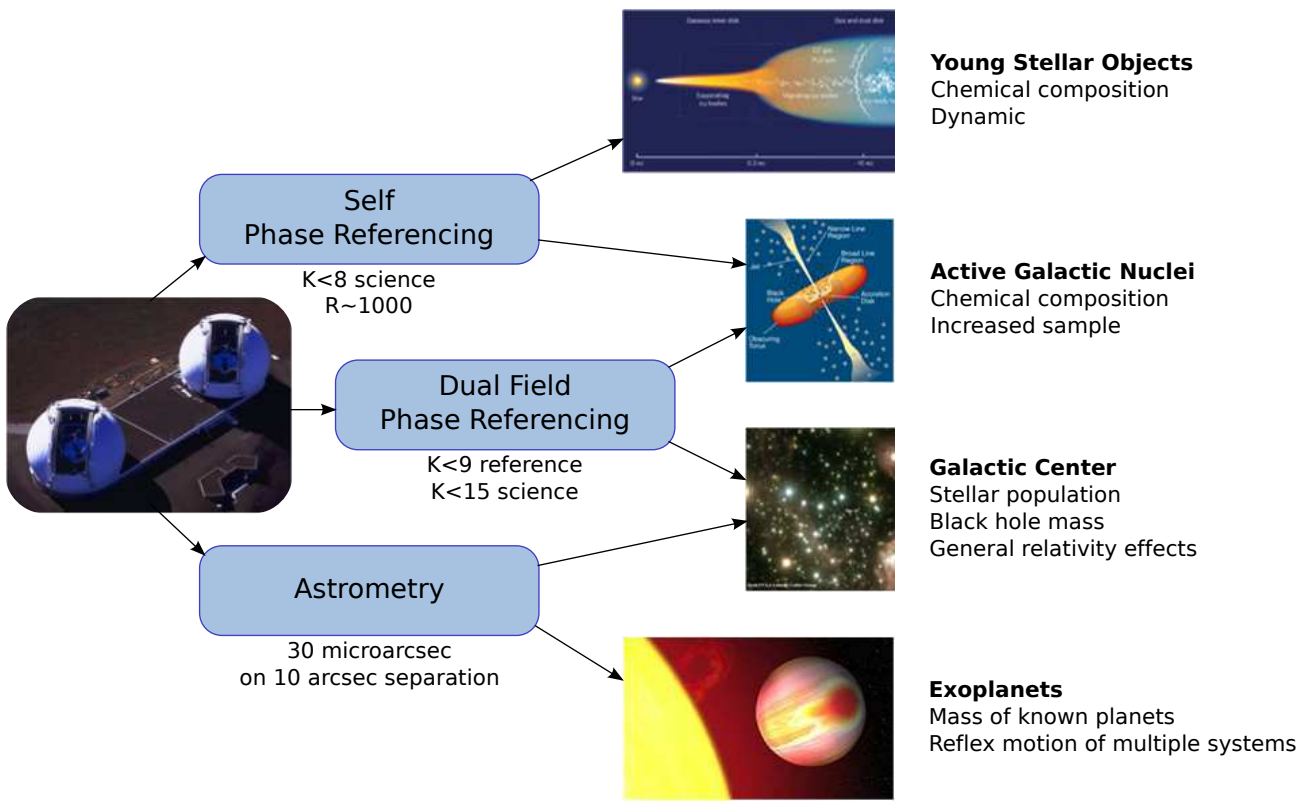}\\
\caption{Summary of the ASTRA project and its science cases.}
\label{Fig:MasterFigure}
\end{figure}

ASTRA, the ASTrometric and phase-Referencing Astronomy upgrade funded by the National Science Foundation, aims at expanding the existing capabilities of the Keck interferometer in three incremental steps. First, a self-phase-referencing mode, based on an on-axis fringe tracker makes possible longer integrations for observations with an $R\sim 2000$ spectral resolution\cite{Woillez+2010,Pott+2010a,Pott+2010b}, aimed at benefiting YSOs observations\cite{Eisner+2010a}. Second, a dual-field phase-referencing mode, based on an off-axis fringe tracker will allow observing fainter objects ($K<15$) with a nearby guide star. This mode will mostly benefit extra-galactic astronomy, increasing by an order of magnitude the number of observable AGN. Third, a narrow angle astrometry mode will measure relative positions with precision of $30 \sim 100$ microarcseconds for objects separated by up to $30$ arcseconds, and further characterize known multi-planet systems. Ultimately, combining the upgraded capabilities of the interferometer with laser guide star adaptive optics on both telescopes will make possible the astrometric monitoring of the inner stars of the galactic center, probing general relativity in the strong field regime. The goals of the project are illustrated in figure \ref{Fig:MasterFigure}, and the main science cases detailed elsewhere in this proceedings\cite{Eisner+2010b}.

With the self-phase referencing mode (SPR, section \ref{Sec:SPR}) in routine operation\cite{Ragland+2010}, this contribution focuses on the status of the dual-field phase referencing (DFPR, section \ref{Sec:DFPR}) and astrometric modes (AST, section \ref{Sec:AST}). For each mode, we will give an overview of the new components added to the interferometer, a summary of which is given in table \ref{Tab:SubsystemsSummary}, as well as a description of the mode and expected or measured performance. In a last section, we present an astrometric calibration scheme, Differential Narrow angle Astrometry (DNA, section \ref{Sec:DNA}), that will let us relax the requirements on the knowledge of the astrometric baseline.

\begin{figure}
\centering
\includegraphics[width=\linewidth]{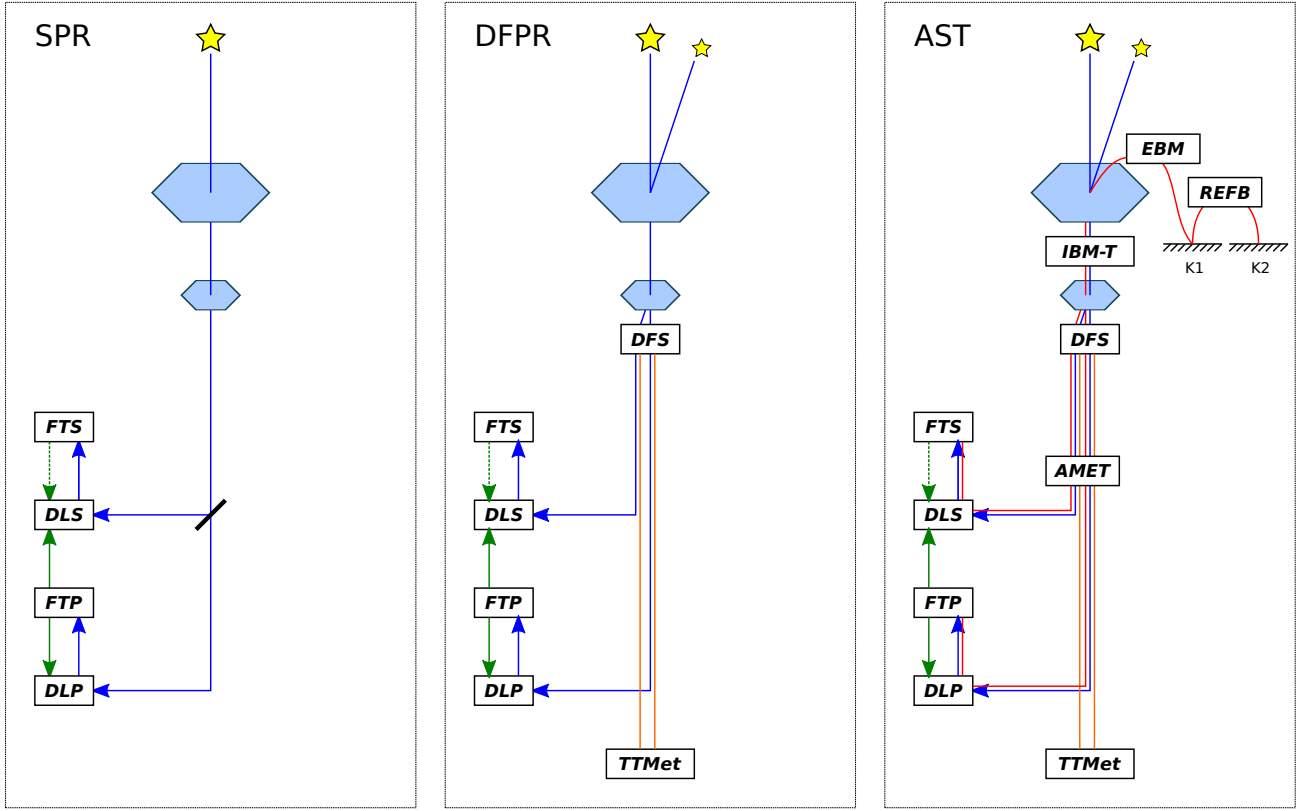}
\caption{
	Architectures of the Self Phase Referencing (SPR, left), Dual Field Phase Referencing (DFPR, center), and Astrometry modes (AST, right). FT-P/S: Primary/Secondary fringe tracker - DLP/S: Primary/Secondary Delay Lines - DFS: Dual Field Subsystem - TTMET: Tip/Tilt Metrology - AMET: Astrometric Metrology - IBM-T: Internal Baseline Monitor Transverse - EBM: External Baseline Monitor - REFB: Reference Baseline
}
\label{Fig:ModesArchitectures}
\end{figure}

\begin{table}
\centering
\begin{tabular}{|c|c|l|c|c|}
\hline
{\bf Mode} & {\bf Mode status} & {\bf Subsystem}                               & {\bf Subsystem status} & {\bf Section}     \\
\hline
\hline
SPR        & operational       & FTS:   Secondary Fringe Tracker Camera        & operational           & \ref{Sec:SPR}     \\
\hline
DFPR       & shared risk       & DFS:   Dual Field Subsystem                   & engineering           & \ref{SSec:DFS}    \\
           &                   & TTMET: Tip Tilt Metrology                     & engineering           & \ref{SSec:TTMET}  \\
\hline
AST        & implementation    & AMET:  Astrometric metrology                  & implementation        & \ref{SSec:AMET}  \\
           &                   & IBMT: Internal Baseline Monitor - Transverse  & design                & \ref{SSec:IBMT}  \\
           &                   & EBM:   External Baseline Monitor              & concept               & \ref{SSec:EBM}   \\
           &                   & REFB:  Reference Baseline                     & concept               & \ref{SSec:REFB}  \\
\hline
\end{tabular}
\vspace{3mm}
\caption{Summary of the ASTRA subsystems, organized by observing mode, with status.}
\label{Tab:SubsystemsSummary}
\end{table}

\section {SELF-PHASE-REFERENCING}\label{Sec:SPR}

Now operational, the first step of the ASTRA project was to deliver an on-axis phase referencing mode called self-phase-referencing (SPR) mode. An extensive description of this mode is being published in Woillez et al. (2010)\cite{Woillez+2010}, along with the first results in Pott et al. (2010)\cite{Pott+2010a,Pott+2010b} and Eisner et al. (2010)\cite{Eisner+2010a}. Observations at $R \sim 2000$ on $K \sim 7$ targets were demonstrated, as well as the detection of differential phase signature of narrow spectral features with an accuracy of $3 mrad$ on $K \sim 5$ targets. This mode required the implementation of an additional fringe tracker camera and beam combiner, to allow different integration times between a fast fringe tracking feed stabilizing the optical path difference for a slow spectrograph. From the perspective of the ASTRA instrumentation program, this mode should be considered as a stepping stone to the dual field mode, that exercises the feed forwarding mechanism from the fringe tracker to the science camera and the handling of slow integrations. The architecture of SPR is illustrated on the left side of Figure \ref{Fig:ModesArchitectures}. For a detailed description of the fringe trackers, developed by the KI project, and extensively used by all the ASTRA modes, see Colavita et al. (2010)\cite{Colavita+2010}.

\section{DUAL-FIELD PHASE-REFERENCING}\label{Sec:DFPR}

The dual field mode can be considered as an improvement on the self phase referencing mode where a nearby guide star is used to stabilize the phase of a fainter object of interest. The fast primary fringe tracker tracks the bright primary reference by sending feed-back targets to its own delay lines. It also sends feed-forward targets to the secondary fringe tracker performing long integration on the fainter object. Existing longitudinal metrology systems (known as coudé metrology) are responsible for removing instrumental phase decorrelations between primary and secondary beam trains, and limit the origin of piston-induced contrast losses to the atmosphere. The configuration of the DFPR mode is illustrated in the central part of Figure \ref{Fig:ModesArchitectures}. The additional subsystems for DFPR are described in section \ref{SSec:DFS} for the Dual Field Subsystem and section \ref{SSec:TTMET} for the Tip/Tilt Metrology. Expected performance and preliminary results from the on-going engineering campaign are presented in section \ref{SSec:PerformanceDFPR}.

\subsection{DFS: Dual Field Subsystem}\label{SSec:DFS}

The dual field subsystem (DFS), located at the focus of the Keck Adaptive Optics (AO) system, is in charge of selecting two different fields and sending them through two separate beam trains (primary/secondary). It relies on an annular mirror located on the optical axis, a few millimeters away from the focus of the Keck Adaptive Optics (Keck AO). A primary on-axis field is transmitted through the central hole of this mirror while a secondary field is reflected. The steerable annular mirror, used in conjunction with another steerable flat mirror, constitute the Off-axis Field Selector (OFS), which can select the secondary field anywhere between an inner radius of 5 arcsec (field of regard inner angle) set by the size of the annular mirror hole, and 30 arcsec (field of regard outer angle). This dual field implementation provides the best transmission at the expense of a somewhat larger field of regard inner angle. The outer angle is limited by the field of view of the Keck AO system initially set to be comparable in size to the isoplanatic patch. After the OFS, both primary and secondary diverging beams are recollimated with off-axis parabolas, and sent through the same primary and secondary beam trains already used by the NULLER and $V^2$ modes. At the motion control level, the OFS implement an algorithm to avoid sending the secondary field through the central hole of the annular mirror, which would otherwise cause unacceptable astrometric metrology breaks (see section \ref{SSec:AMET}). At a higher level, the OFS are driven by the interferometer sequencer which automatically perform the acquisition of the off-axis target, based on its offset with respect to the on-axis target and the orientation of the field obtained form the telescope control system.  Finally, the ASTRA DFS is an interface between Keck AO and the beam trains completely different from the one used for the NULLER configuration, which uses a collimated output of the AO system directly. A reconfiguration of this interface is therefore needed between ASTRA and NULLER; whereas the $V^2$ modes can, as of now, use any of the two modes.

The operation and performance of the DFS has been demonstrated on sky. Initial off-axis acquisition performed while the AO loops are being closed do not add any overhead to the acquisition process. Target swaps between primary and secondary are performed in under $\sim 15$ seconds and will provide a high duty cycle to the astrometric metrology calibration. 

\subsection{TTMET: Tip Tilt Metrology}\label{SSec:TTMET}

With a faint object in one of the two beam train, the angle tracker traditionally running at 80 Hz on bright objects and operating at J or H band, has to be slowed down accordingly. Then, the achieved correction bandwidth can only address very slow drifts of thermal or alignment origin, leaving tip/tilt vibrations and beam train turbulence uncorrected. A tip/tilt metrology system has been developed to deal with these high frequencies. This system is composed of visible (690 nm) light sources injected collimated in the beam train after the DFS, and sent to 2D position sensitive sensors, located in the interferometric basement past the pick-up dichroics to the beam combiners. The output of the position sensors, along with the output of the angle tracker camera, jointly feed the fast tip/tilt mirrors in an inner/outer loop scheme. The fast TTMET loop typically run at 2 kHz, while the angle tracker can integrate on second time scales. For objects extremely faint in H band, the infrared angle tracker can be disabled completely and the system relies entirely on the tip/tilt metrology. This is a yet untested operational mode for TTMET that will require a thorough investigation of the pointing accuracy of the AO and OFS. Limited to only one angle tracker camera, neutral densities have to be inserted in the beam to avoid saturation of the bright object on the slow camera.

This system has been completely installed on the summit, and basic functionalities demonstrated. A complete performance characterization is still needed, especially a comparison of the injection stability on the fringe trackers between the angle tracker and TTMET only configurations.

\subsection{DFPR mode: Expected performance and preliminary results}\label{SSec:PerformanceDFPR}

The first parameter that characterizes DFPR is the limiting magnitude of the fast fringe tracker. Despite a different interface with AO, the DFPR mode has a throughput identical to the traditional $V^2$ setup. The advertised $K \sim 10$ $V^2$ limiting magnitude obtained at a fringe rate of 200 Hz, offset by $-1$ magnitude to achieve a stable fringe lock, gives a DFPR limiting magnitude of $K \sim 9$, confirmed on sky.

\begin{figure}
\centering
\includegraphics[width=\linewidth]{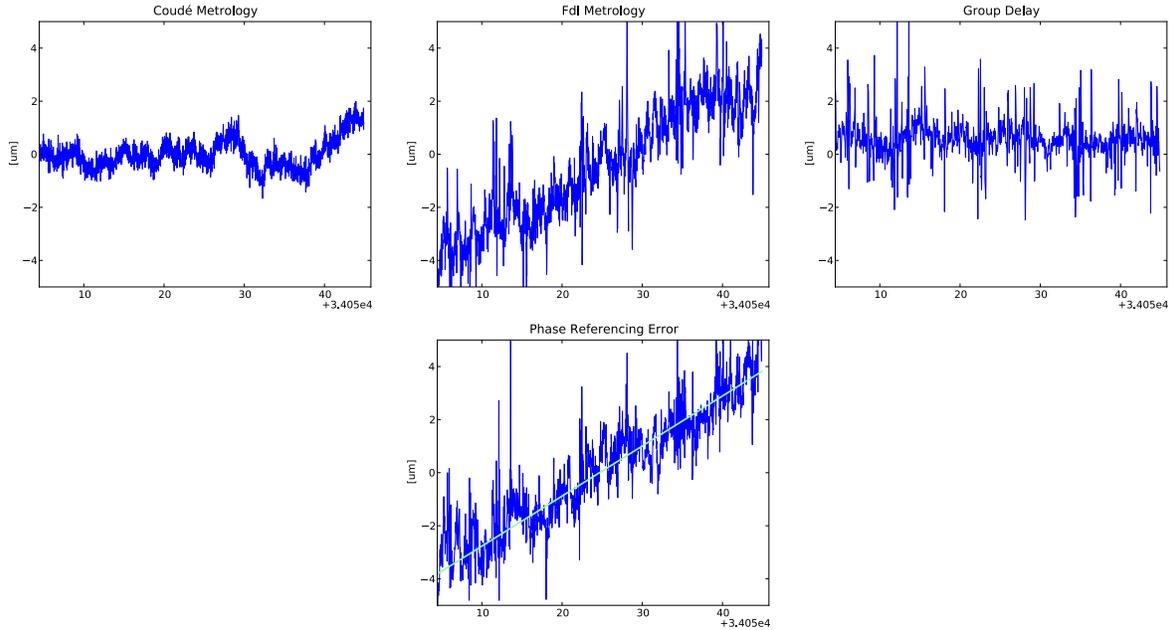}
\caption{Phase referencing performance. Top-Left: Coudé metrology differential OPD - Top-Center: Fast delay lines differential OPD - Top-Right: Differential group delay. - Bottom: Total differential OPD (combination of the top quantities) versus predicted differential OPD for the HD16410/HD16411 pair.}
\label{Fig:DifferentialPiston}
\end{figure}

Thus far, we have only demonstrated parallel fringes on a bright-bright pair where both fringe trackers were independently running at fast rates. The data collected, and presented in Figure \ref{Fig:DifferentialPiston}, gives an idea of the current phase referencing performance. From the low frequency perspective there is no issue: the fringe positions follow the prediction for the pair separation and is not a limit to the integration time. From the high frequency perspective, we still have a unacceptably high level of uncorrected vibrations. This is the main issue we need to address for DFPR.

\section{ASTROMETRY}\label{Sec:AST}

In order to make the DPFR mode of ASTRA capable of narrow-angle astrometry measurement, metrology systems have to be added to the interferometer to cover the entire differential OPD involved in the narrow angle measurement. First to be implemented, an astrometric metrology (AMET, section \ref{SSec:AMET}) measures the differential OPD from the primary and secondary beam combiner beam splitters up to a common (in this primary versus secondary sense) corner cube located in front each AO deformable mirror, inside the telescope central obscuration of each telescope. The astrometric metrology also materializes the astrometric baseline defined as the vector joining the apexes of the metrology corner cubes, expressed in the primary space of each telescope (in front of the primary mirror). The metrology of the astrometric baseline itself is achieved through the combination of an internal baseline monitor - transverse (IBMT, section \ref{SSec:IBMT}) responsible for monitoring the transverse conjugation of each metrology corner cube from inside the AO system to primary space, and an external baseline monitor (EBM, section \ref{SSec:EBM}) responsible for monitoring the position of the metrology corner cube conjugate in primary space with respect to a ground reference frame in each telescope. The Reference Baseline (REFB, section \ref{SSec:REFB}) is the last component of the astrometric baseline metrology. It links together the ground reference frames of the two telescopes. The overall organization of the astrometric metrology subsystems is illustrated on the right side of Figure \ref{Fig:ModesArchitectures}.

Currently, AMET and IBMT are the only two subsystems being implemented: they have been identified as addressing the largest error terms involved in the astrometric measurement. They will provide a basic astrometric capability together with a Differential Narrow Angle observing technique (DNA, section \ref{Sec:DNA}), which reduces the baseline knowledge requirement. EBM and REFB are kept at the concept level for now. A decision on whether to implement them will be made once this basic astrometry mode is operational, sometime in 2011.

\subsection{AMET: Astrometric Metrology}\label{SSec:AMET}

\begin{figure}
\centering
\includegraphics[width=0.85\linewidth]{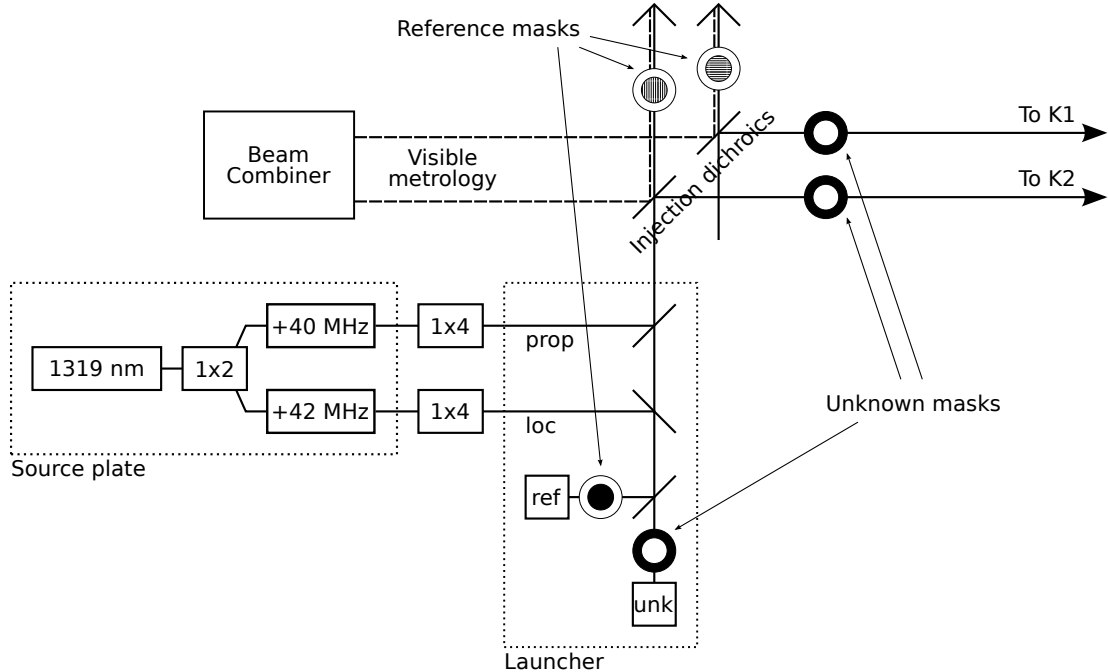}\\
\caption{Overview of the Astrometric Metrology (AMET). See details in text (section \ref{SSec:AMET}).}
\label{Fig:AMET}
\end{figure}

The main component of the astrometric metrology is an 1319 nm heterodyne metrology system, very similar in concept to the metrology gauge\cite{Gursel2003} designed for the Differential Phase instrument\cite{Akeson+2000} (itself inspired from an early SIM metrology system\cite{Ames+2003}), where orthogonal masks are used to isolate two metrology propagations with a low level of self-interference and cross-talk. In the ASTRA implementation, an annular mask selects the propagation through the beam train up to the primary/secondary corner cube in the AO systems (conveniently letting the star light through as well), while a central mask selects the annular propagation to a reference corner cube located on the opposite side of the injection beam splitter. Each of the four metrology launchers will measure the difference between the path from the injection dichroic to the metrology corner cube, and the path from the dichroic and the reference corner cube. To complete the beam train coverage, another existing metrology system based on a single stabilized HeNe laser source, measures the on-axis optical path difference form the beam combiner beam splitters to the reference corner cubes. This internal metrology system was developed for the NULLER but is not used currently. The propagation selection for this visible system is based on selecting orthogonal polarizations with polarizers. For the reference corner cube to work with both system, it is composed of an on-axis inner mask that blocks the infrared system, and also selects one single linear polarization for the visible system. Figure \ref{Fig:AMET} gives a summary of the AMET setup. The detection electronics for the infrared system is an exact replica of the systems in use for the other metrology systems, except for the actual detector adapted to infrared wavelengths.

To reach the sub-micron differential OPD accuracy requirement, the infrared laser wavelength should be stabilized with a relative accuracy better than the $10^{-6}$ goal for the astrometric relative accuracy. Instead, we have chosen to relax this requirement by operating the infrared system at a differential OPD close to zero. This can be achieved by adjusting the position of the reference corner cubes to match the differential OPD in the delay lines. In practice, this is equivalent to a transfer of the differential OPD measurement to the visible metrology system, which already meet the wavelength stability requirement without need of an additional stabilization.

Both visible and infrared systems are relative metrology systems: they only provide an accurate variation of the optical path relative to an offset, which can be assumed fixed  as long as the metrology systems stay uninterrupted and the system stable. The determination of this offset is achieved through a swap of the observed pair of targets between the primary and secondary beams. This operation corresponds to a sign change for the separation vector, which makes the half sum of differential OPD give directly the metrology offset. This swap operation is the basis of an astrometric measurement for the ASTRA implementation.

The lab integration of the AMET subsystem has just started. The installation on the summit is expected to happen through the summer of 2010, for on-sky tests scheduled in the fall. 

\subsection{IBMT: Internal Baseline Monitor - Transverse}\label{SSec:IBMT}

\begin{figure}
\includegraphics[width=0.6\linewidth]{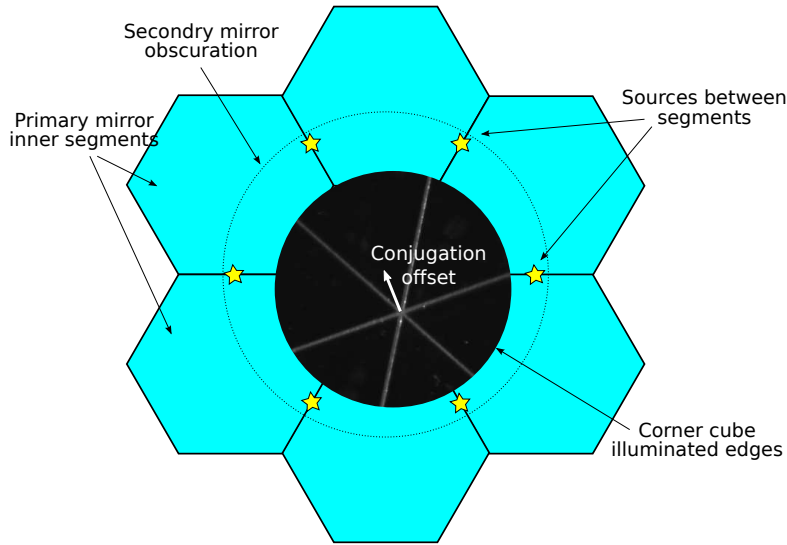}
\centering
\caption{Overview of the Internal Baseline Monitor - Transverse (IBMT). See details in text (section \ref{SSec:IBMT}).}
\label{Fig:IBMT}
\end{figure}

The metrology corner cube located in front of the AO deformable mirrors define the astrometric baseline. This baseline needs to be precisely known in primary space, expressed in the earth reference frame. The Internal Baseline Monitor takes care of monitoring the conjugation from inside the AO system to primary space. Since the vector separation between the astrometric pair is orthogonal to the direction of the pair, only the component transverse to the pointing direction needs to be monitored. The Internal Baseline Monitor Transverse (IBMT) is designed as an imaging camera located inside the AO system and conjugated to the primary mirror and deformable mirror. Sources are installed in the gaps between the first row of the primary mirror segments. Imaged by this camera, the position of these sources with respect to the illuminated edges of the metrology corner cube represents the conjugation of the astrometric baseline in primary space, with respect to the primary mirror itself. The conjugation vector measured on second timescales, will be available immediately to correct the differential OPD target sent to the delay lines, and will be archived for post-processing of the astrometric observables.   

Figure \ref{Fig:IBMT} gives an illustration of the IBMT setup, along with an actual image of the metrology corner cube edges illuminated in the lab. The planned IBMT implementation uses a 2kx2k camera with a 2~mm/pix primary space plate scale. A primary space accuracy of $100 \mu m$ corresponds to a reasonable pixel/20 position accuracy requirement for the metrology corner cube edges and primary space sources.

\subsection{EBM: External Baseline Monitor}\label{SSec:EBM}

The external baseline monitor would be in charge of measuring the position of the primary mirror with respect to the ground and therefore complete the baseline monitoring requirement that IBMT only partially fulfills. The concept for this subsystem is based on off-the-shelf 3D laser trackers. Located on the ground in each telescope they would be tracking the motion of corner cubes tied to the primary mirror. Since the primary mirror itself is hardly visible from the telescope floor, and in order to properly connect EBM to IBMT, the reference sources currently located at the primary mirror would have to move to the sides of the secondary mirror (still in primary space), and be tied to corner cube retro reflectors that laser trackers are capable of measuring. A network of reference corner cubes anchored to the ground would then provide an earth bound reference frame for each telescope. The contributions from EBM to the estimation of the astrometric baseline will be added to those from IBMT.

\subsection{REFB: Reference Baseline}\label{SSec:REFB}

The final step to complete the astrometric baseline determination is to measure the vector joining the ground reference frames of each telescope. This vector, or reference baseline (REFB), can be measured following two alternative procedures: directly in the earth reference frame using surveying techniques, or using a wide angle baseline calibration. While the first method relies entirely on the tool used for the survey, the second method might require the addition of another metrology system to the telescopes. Whereas a narrow angle astrometry measurement requires the IBM to deal with the transverse conjugation of the metrology corner cube to primary space, a wide angle measurement requires a control in the longitudinal direction: the optical path from the metrology corner cube to the primary space. This additional metrology system would have to measure the optical path through the telescope. Keep in mind that the accuracy of the determination of the wide angle baseline relies entirely on the knowledge of the optical path through the interferometer.

\section{DIFFERENTIAL NARROW ANGLE ASTROMETRY}\label{Sec:DNA}

\begin{figure}
\centering
\includegraphics[width=0.75\linewidth]{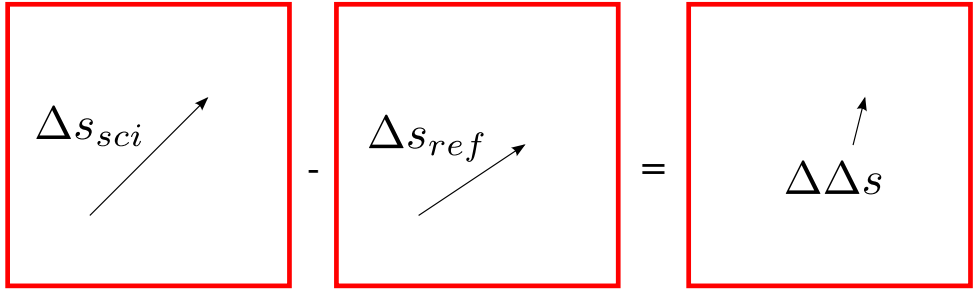}
\caption{DNA astrometry. A reference pair (middle) is found to approximately match the separation of the science pair (left), in order to have a smaller combined separation (right).}
\label{Fig:DNA}
\end{figure}

Without a full implementation of the astrometric baseline metrology, we have investigated a way to reduce the requirement on the absolute knowledge of the astrometric baseline (which EBM and REFB would contribute to, if implemented). The concept is to use a reference pair $\Delta{s}_{ref}$ of similar projected separation and close to the pair of astrometric interest $\Delta{s}_{sci}$, so that the differential separation $\Delta\Delta{s} = \Delta{s}_{sci} - \Delta{s}_{ref}$ becomes smaller (illustration in Figure \ref{Fig:DNA}). For the expolanet case, performing such a reference only adds the proper motion of the reference pair to the proper motion of the exoplanet pair. For the galactic center case, we are however losing an important observable: the absolute scale of the astrometric measurement. Finding a reference pair matching a 10 arcsec science pair to within 1 arcsec, reduces the absolute baseline knowledge requirement by a factor of 10. However, in order to be able to combine both reference and science measurements, any relative change to the baseline must still be accounted for at the initial requirement level. This statement can be supported with the two following astrometric observations, where the second observation is differential with respect to the the first observation.
\[
	\left\{
		\begin{array}{c}
B\cdot\Delta{s} + \Delta{OPD} = 0 \\
(B+\Delta{B})\cdot(\Delta{s}+\Delta\Delta{s}) + (\Delta{OPD}+\Delta\Delta{OPD}) = 0 \\
		\end{array}
	\right.
\]
From the first observation to the second, the baseline has changed by $\Delta{B}$, the separation to be measured by $\Delta\Delta{s}$, and the differential OPD by $\Delta\Delta{OPD}$. Subtracting the first observation from the second, keeping the dominant terms, and differentiating for the error analysis, we get:
\[ \delta{B}\cdot\Delta\Delta{s} + B\cdot\delta\Delta\Delta{s} + \delta\Delta{B}\cdot\Delta{s} + \Delta{B}\cdot\delta\Delta{s} + \delta\Delta\Delta{OPD} = 0 \]

The second term is the one that drives the requirements, containing $\delta\Delta\Delta{s}$ the error on the differential narrow angle astrometry. Its comparison to the other terms sets the requirements for DNA astrometry.
\[
\begin{tabular}{rl}
   Absolute baseline accuracy $\delta{B}$:                                      & $ \delta{B} < B \frac{\delta\Delta\Delta{s}}{\Delta\Delta{s}} $ \\
   Relative baseline accuracy $\delta\Delta{B}$:                                & $ \delta\Delta{B} < B \frac{\delta\Delta\Delta{s}}{\Delta{s}} $ \\
   Relative baseline motion $\Delta{B}$:                                        & $ \Delta{B} < B \frac{\delta\Delta\Delta{s}}{\delta\Delta{s}} $ \\
   Double differential optical path difference error $\delta\Delta\Delta{OPD}$: & $ \delta\Delta\Delta{OPD} < B\delta\Delta\Delta{s} $
\end{tabular}
\]
This analysis does not take into account the baseline/separation geometry which could be added in a similar fashion with additional requirements and without changes to the conclusion.

Table \ref{Tab:DNAErrorBudget} presents these constraints in term of the Exoplanet and Galactic Center science cases. We have assumed an initial separation accuracy $\delta\Delta{s}$ of $100 mas$. We also present the requirement on the individual differential optical path differences ($\delta\Delta{OPD}$, assumed independent between the science and reference pairs) rather than on the difference itself ($\delta\Delta\Delta{OPD}$) to illustrate that we now have a $\sqrt{2}$ tighter requirement compared to the classical narrow angle astrometry case.

To the expense of a lower observing efficiency (with the addition of the reference pair, any astrometric measurement now takes twice as much time), DNA astrometry should let us perform initial astrometric measurements on the Galactic Cetner and a few Exolplanets, without the implementation of all the astrometric baseline metrology systems.

\begin{table}
\centering
\begin{tabular}{|c|c|c|c|c|c|c|c|c|}
\hline
        &              \multicolumn{3}{|c|}{Assumptions}      &       Target error      &                       \multicolumn{4}{|c|}{Constraints}             \\
  Case  & $\Delta{s}$ & $\delta\Delta{s}$ & $\Delta\Delta{s}$ & $\delta\Delta\Delta{s}$ & $\delta{B}$ & $\delta\Delta{B}$ & $\Delta{B}$ & $\delta\Delta{OPD}$ \\
\hline
\hline
  ExoP  &   $20''$    &     $100 m''$     &      $1''$        &       $50 \mu''$        & $< 4.25 mm$ &  $< 212 \mu m$    & $< 42.5 mm$ &    $< 14.6 nm$      \\
   GC   &   $ 6''$    &     $100 m''$     &      $1''$        &       $30 \mu''$        & $< 2.55 mm$ &  $< 425 \mu m$    & $< 25.5 mm$ &    $<  8.7 nm$      \\
\hline
\end{tabular}
\caption{DNA astrometry error budget (without baseline geometry contribution)}
\label{Tab:DNAErrorBudget}
\end{table}

\acknowledgments     

The W. M. Keck Observatory is operated as a scientific partnership among the California Institute of Technology, the University of California, and the National Aeronautics and Space Administration. The Observatory was made possible by the generous financial support of the W. M. Keck Foundation. The authors wish to recognize and acknowledge the very significant cultural role and reverence that the summit of Mauna Kea has always had within the indigenous Hawaiian community. We are most fortunate to have the opportunity to conduct observations from this mountain. We would like to acknowledge the Keck Interferometer project funded by the National Aeronautics and Space Administration (NASA) and developed and operated by the Jet Propulsion Laboratory, the W. M. Keck Observatory and the NASA Exoplanet Science Institute. This material is based upon work supported by the National Science Foundation under Grant No. AST-0619965.

\bibliography{document}
\bibliographystyle{spiebib}

\end{document}